\documentclass[twocolumn,twoside,slac_two]{revtex4}
\usepackage{graphicx}
\usepackage{fancyhdr}
\def \b{{\cal B}}
\def \bea{\begin{eqnarray}}
\def \beq{\begin{equation}}
\def \ca{{\cal A}}
\def \eea{\end{eqnarray}}
\def \eeq{\end{equation}}
\def \od{\overline{D}^0}
\def \ok{\overline{K}^0}
\def \s{\sqrt{2}}
\def \st{\sqrt{3}}
\def \sx{\sqrt{6}}
\def \ta{\tilde{A}}
\def \tc{\tilde{C}}
\def \te{\tilde{E}}
\def \ttl{\tilde{T}}

\begin{document}

\title{Flavor Symmetry and Charm Decays}

%

\author{Bhubanjyoti Bhattacharya and Jonathan L. Rosner}
\affiliation{Enrico Fermi Institute and Department of Physics, University
of Chicago, Chicago, IL 60637}

\begin{abstract}
A wealth of new data in charmed particle decays allows the testing of flavor
symmetry and the extraction of key amplitudes.  Information on relative
strong phases is obtained.
\end{abstract}

\maketitle

\thispagestyle{fancy}


\section{Introduction}

The application of flavor symmetries, notably SU(3), to charmed particle
decays can shed light on some fundamental questions.  Often it is useful
to know the strong phases of amplitudes in these decays.  For example, the
relative strong phase in $D^0 \to K^- \pi^+$ and $\od \to K^- \pi^+$ is
important in interpreting decays of $B$ mesons to $D^0 X$ and $\od X$
\cite{Gronau:2001nr,ADS}.  Such strong phases are non-negligible even in $B$
decays to pairs of pseudoscalar mesons ($P$) despite some perturbative QCD
expectations to the contrary, and can be even more important in
$D \to PP$ decays.  In the present report we shall illustrate the extraction
of strong phases from charmed particle decays using SU(3) flavor symmetry,
primarily the U-spin symmetry involving the interchange of $s$ and $d$ quarks.

We begin in Section 2 by discussing the overall diagrammatic approach to
flavor symmetry.  In Section 3 we treat Cabibbo-favored decays, turning to
singly-Cabibbo-suppressed decays in Section 4 and doubly-Cabibbo-suppressed
decays in Section 5.  We note specific applications to $D^0$ and $\od$
decays to $K^- \pi^+$ in Section 6, mention some other theoretical approaches
in Section 7, and conclude in Section 8.

\section{Diagrammatic amplitude expansion}

\begin{figure*}[t]
\mbox{\includegraphics[width=0.46\textwidth]{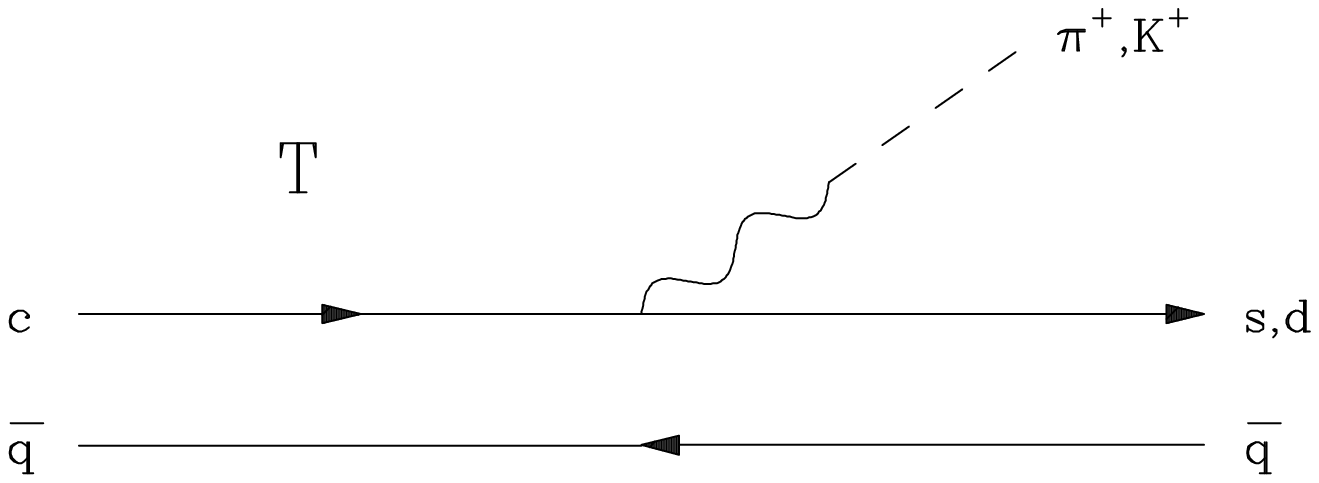} \hskip 0.3in
      \includegraphics[width=0.46\textwidth]{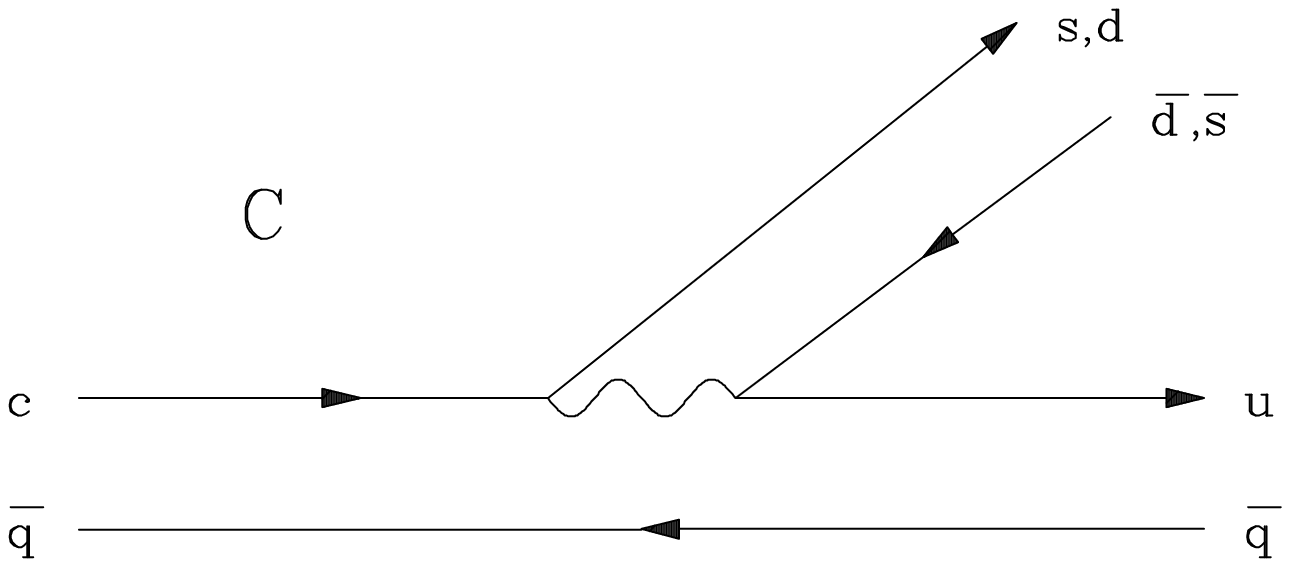}}
\vskip 0.3in
\mbox{\includegraphics[width=0.46\textwidth]{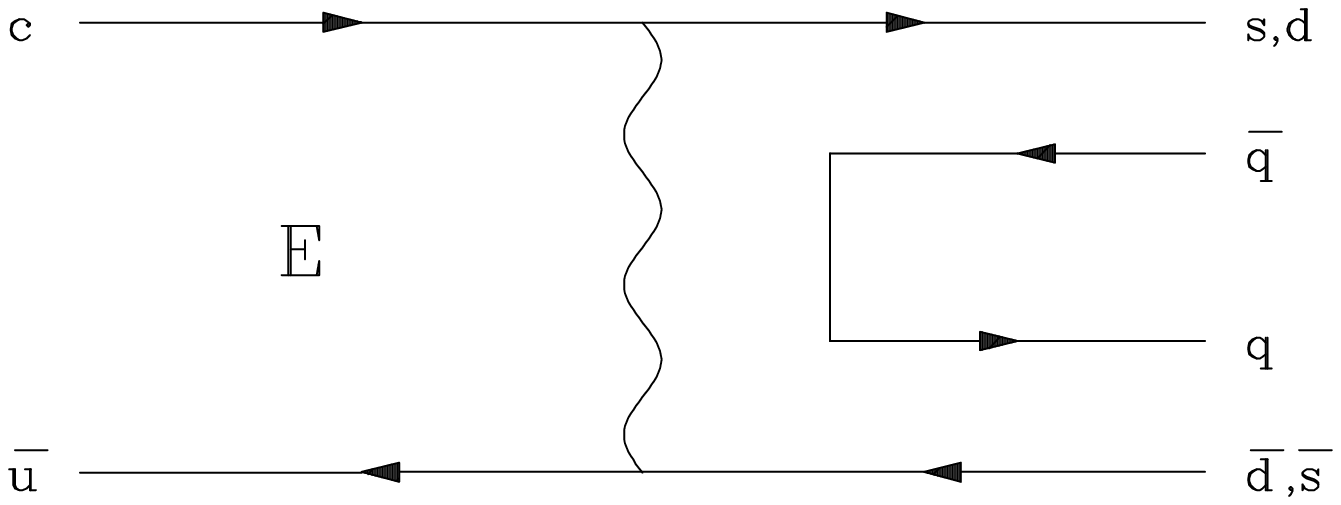} \hskip 0.3in
      \includegraphics[width=0.46\textwidth]{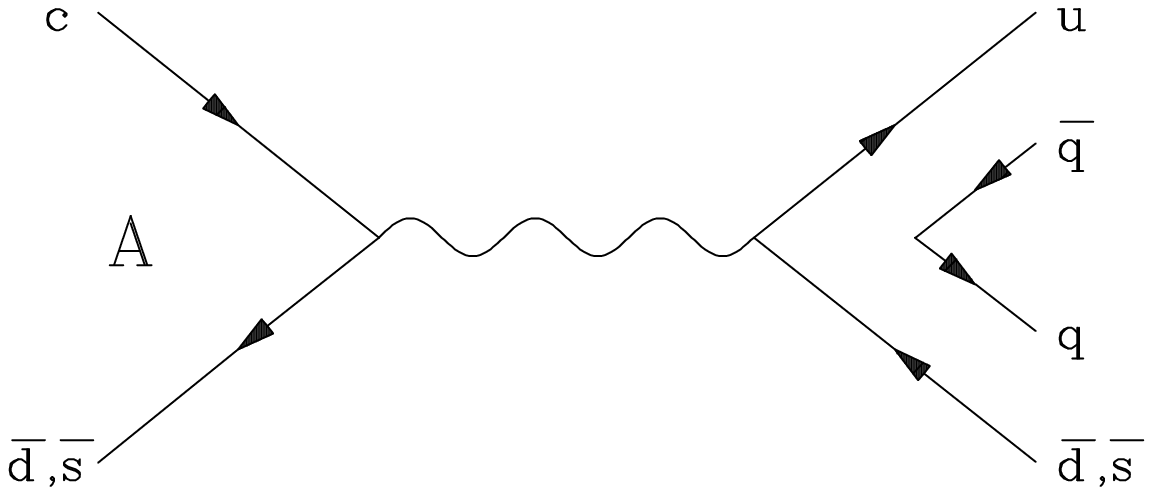}}

\caption{Flavor topologies for describing charm decays.  $T$: color-favored
tree; $C$: color-suppressed tree; $E$ exchange; $A$: annihilation.
\label{fig:TCEA}}
\end{figure*}

We use a flavor-topology language for charmed particle decays first
introduced by Chau and Cheng \cite{Chau:1983,Chau:1986}.  These topologies,
corresponding to linear combinations of SU(3)-invariant amplitudes, are
illustrated in Fig.\ \ref{fig:TCEA}.  Cabibbo-favored
(CF) amplitudes, proportional to the product $V_{us} V^*_{cs}$ of
Cabibbo-Kobayashi-Maskawa (CKM) factors, will be denoted by unprimed
quantities; singly-Cabibbo-suppressed amplitudes proportional to $V_{us}
V^*_{cs}$ or $V_{ud} V^*_{cd}$ will be denoted by primed quantities; and
doubly-Cabibbo-suppressed quantities proportional to $V_{us} V^*_{cd}$ will
be denoted by amplitudes with a tilde.  The relative hierarchy of
these amplitudes is $1:\lambda:-\lambda:-\lambda^2$, where $\lambda =
\tan \theta_C = 0.232 \pm 0.002$ \cite{Yao:2006px}.  Here $\theta_C$ is the
Cabibbo angle.

\section{Cabibbo-favored decays}

A detailed discussion of amplitudes and their relative phases for
Cabibbo-favored charm decays was given in Ref.\ \cite{Rosner:1999}.  The main
conclusions of that analysis were large relative phases of the $C$ and $E$
amplitudes relative to the dominant $T$ term, and an approximate relation
$A \simeq -E$.  The present updated data confirm these results.

In Table \ref{tab:CF} we show the results of extracting amplitudes ${\cal A} =
M_D[8 \pi {\cal B} \hbar/(p^* \tau)]^{1/2}$ from the branching ratios
${\cal B}$ and lifetimes $\tau$, all from Ref.\ \cite{Yao:2006px}
unless otherwise noted.  Here $M_D$ is the mass of the decaying
charmed particle, and $p^*$ is the final c.m.\ 3-momentum.

The extracted amplitudes, with $T$ defined to be real, are, in units of
$10^{-6}$ GeV:
\bea
T & = &  2.71 + 0~  i~; \\
C & = & -1.77 - 1.01i~;~~\delta(CT) = -150^\circ~; \\
E & = & -0.71 + 1.49i~;~~\delta(ET) = 115^\circ~; \\
A & = &  0.57 - 1.30i~;~~\delta(AT) = - 66^\circ~.
\eea
These values update (and are consistent with) those quoted with less
precision in Ref.\ \cite{Rosner:1999}.  New (mainly lower)
preliminary branching ratios for many $D_s$ decays reported at this
Conference \cite{Ryd:2007} will change some of the results
slightly once they are incorporated into averages.

\begin{table*}[t]
\caption{Branching ratios, amplitudes, and graphical representations
for Cabibbo-favored charmed particle decays.
\label{tab:CF}}

\begin{center}
\begin{tabular}{|c|c|c|c|c|c|} \hline
Meson &    Decay    &     $\b$      & $p^*$ &    $|\ca|$    & Rep. \\
      &    mode     &     (\%)      & (MeV) &($10^{-6}$ GeV)&      \\ \hline
$D^0$ & $K^- \pi^+$ & 3.82$\pm$0.07 &  861  & 2.49$\pm$0.03 & $T+E$ \\
      & $\ok \pi^0$ & 2.26$\pm$0.24 &  860  & 1.92$\pm$0.06 & $(C-E)/\s$ \\
      & $\ok \eta$  & 0.76$\pm$0.12 &  772  & 1.18$\pm$0.05 & $C/\st$ \\
      & $\ok \eta'$ & 1.82$\pm$0.28 &  565  & 2.13$\pm$0.09 & $-(C+3E)/\sx$ \\
\hline
$D^+$ & $\ok \pi^+$ & 2.94$\pm$0.12 &  863  & 1.38$\pm$0.02 & $C+T$ \\ \hline
$D_s^+$& $\ok K^+$  & 4.50$\pm$0.80 &  850  & 2.60$\pm$0.25 & $C+A$ \\
      & $\pi^+\eta$ & 2.16$\pm$0.30 &  902  & 1.75$\pm$0.14 & $(T-2A)/\st$ \\
      & $\pi^+\eta'$& 4.80$\pm$0.60 &  743  & 2.88$\pm$0.20 &$2(T+A)/\sx$ \\
\hline
\end{tabular}
\end{center}
\end{table*}

\begin{figure}
\includegraphics[width=80mm]{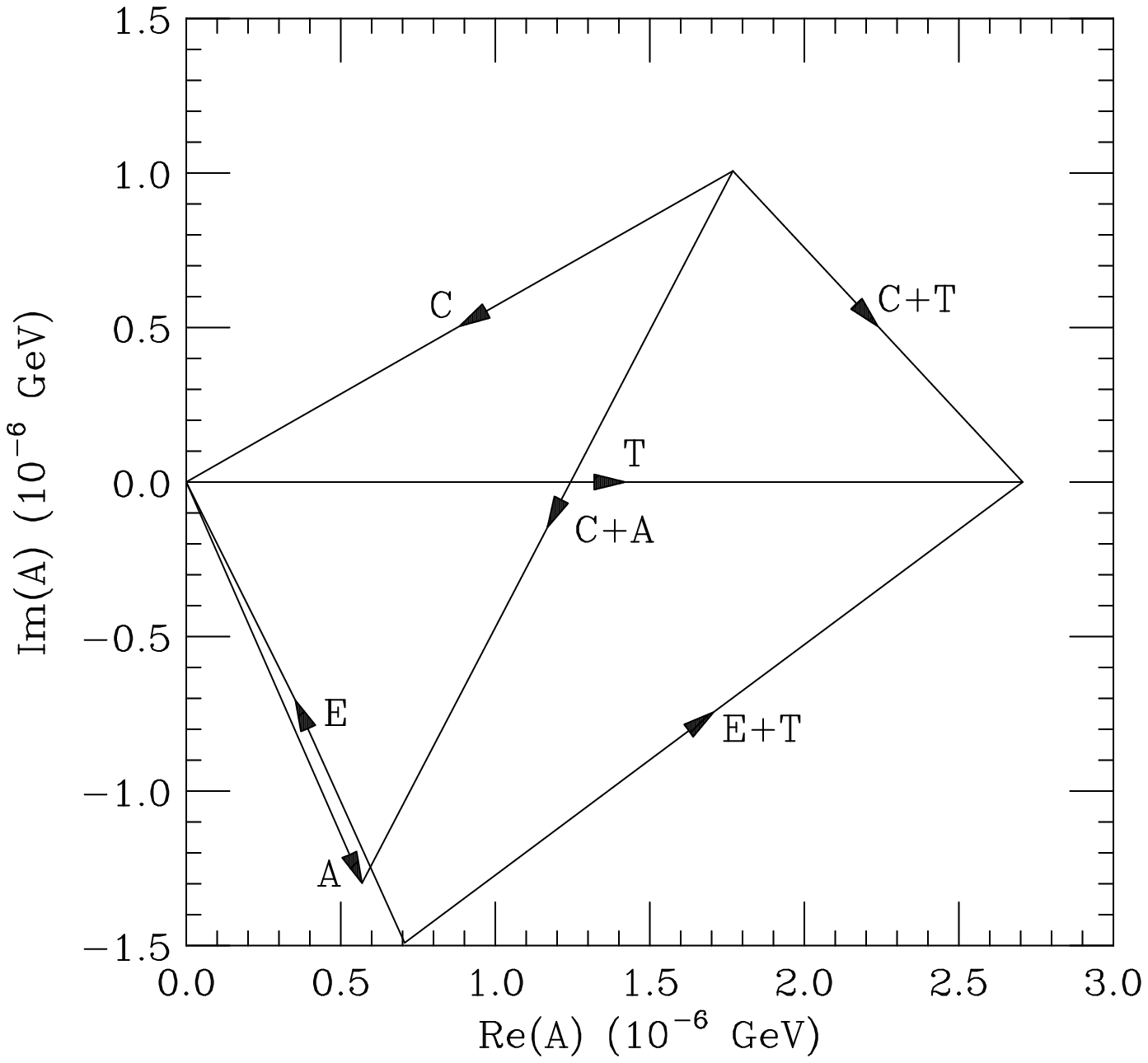}
\caption{Construction of Cabibbo-favored amplitudes from observed processes.
Here the sides $C+T$, $C+A$, and $E+T$ correspond to measured processes;
the magnitudes of other amplitudes listed in Table \ref{tab:CF} are also
needed to specify the reduced amplitudes $T$, $C$, $E$, and $A$.
\label{fig:cf}}
\end{figure}

The Cabibbo-favored amplitudes are shown on an Argand diagram in
Fig.\ \ref{fig:cf}.  Here $A$ was extracted from $D_s \to \pi^+ \eta$ and
$D_s \to \pi^+ \eta'$; the amplitude $\ca$ for $D_s \to \ok K^+$ is then
predicted to be $2.60 \times 10^{-6}$ GeV vs.\ $(2.60 \pm 0.25) \times
10^{-6}$ GeV observed.  Note the importance of the $E$ and $A \simeq -E$
amplitudes.

\section{Singly-Cabibbo-suppressed decays}

\subsection{SCS decays involving pions and kaons}

We show in Table \ref{tab:scskpi} the branching ratios, amplitudes, and
representations in terms of reduced amplitudes for singly-Cabibbo-suppressed
(SCS) charm decays involving pions and kaons.  The ratio of primed (SCS)
to unprimed (CF) amplitudes is expected to be $\tan \theta_C \simeq 0.232$.

\begin{table*}
\caption{Branching ratios, amplitudes, and decomposition in terms of reduced
amplitudes for singly-Cabibbo-suppressed (SCS) charm decays involving pions and
kaons.
\label{tab:scskpi}}
\begin{center}
\begin{tabular}{|c|c|c|c|c|c|} \hline
Meson &    Decay    &     $\b$      & $p^*$ &    $|\ca|$    & Rep. \\
      &    mode     &  $(10^{-3})$  & (MeV) &($10^{-7}$ GeV)&      \\ \hline
$D^0$ &$\pi^+ \pi^-$& 1.37$\pm$0.03 &  922  & 4.57$\pm$0.05 & $-(T'+E')$ \\
      &$\pi^0 \pi^0$& 0.79$\pm$0.08 &  923  & 3.46$\pm$0.18 & $-(C'-E')/\s$ \\
      &  $K^+ K^-$  & 3.85$\pm$0.09 &  791  & 8.26$\pm$0.10 & $(T'+E')$ \\
      &  $K^0 \ok$  & 0.72$\pm$0.14 &  789  & 3.58$\pm$0.35 &    0   \\ \hline
$D^+$ &$\pi^+ \pi^0$& 1.28$\pm$0.08 &  925  & 2.77$\pm$0.09 & $-(T'+C')/\s$ \\
      &  $K^+ \ok$  & 5.90$\pm$0.38 &  793  & 6.43$\pm$0.21 & $T'-A'$ \\ \hline
$D_s^+$&$\pi^+ K^0$ & 2.46$\pm$0.40 &  916  & 5.87$\pm$0.48 & $-(T'-A')$ \\
      & $\pi^0 K^+$ & 0.75$\pm$0.28 &  917  & 3.24$\pm$0.60 &$-(C'+A')/\s$ \\
\hline
\end{tabular}
\end{center}
\end{table*}

The deviations from flavor SU(3) implicit in Table \ref{tab:scskpi} are well
known.  We shall discuss amplitudes in units of $10^{-7}$ GeV.  If one rescales
the CF amplitudes by the factor of $\tan \theta_C$, one predicts
$|\ca(D^0 \to \pi^+ \pi^-)| = |\ca(D^0 \to K^+ K^-)| = 5.78$, to be compared
with a smaller observed value for $\pi^+ \pi^-$ and a larger observed value (by
a factor of $\s$) for $K^+ K^-$.  One can account for some of this discrepancy
via the ratios of decay constants $f_K/f_\pi = 1.22$ and form factors
$f_+(D \to K)/f_+(D \to \pi) > 1$.  Furthermore, one predicts $|\ca(D^0 \to
\pi^0 \pi^0)| = 4.45$ (larger than observed) and $|\ca(D^+ \to \pi^+ \pi^0)|
= 2.25$ (smaller than observed), which means that the $\pi \pi$ isospin
triangle [associated with the fact that there are two independent amplitudes
with $I=(0,2)$ for three decays] has a different shape from that predicted
by rescaling the CF amplitudes.  One predicts $|\ca(D^+ \to K^+ \ok)| =
|\ca(D_s \to \pi^+ K^0)| = 5.79$; experimental values are (11\%,1\%) higher.
The decay $D^0 \to K^0 \ok$ is forbidden by SU(3); the branching ratio of
$2 {\cal B}(D^0 \to K_S^0 K_S^0) = (2.98 \pm 0.68 \pm 0.30 \pm 0.60) \times
10^{-4}$ reported by CLEO \cite{Ryd:2007}
is more than a factor of two below that quoted in Table \ref{tab:scskpi}
(based on the average in Ref.\ \cite{Yao:2006px}) and so does offer some
evidence for the expected suppression.

\subsection{SCS decays involving $\eta,\eta'$}

The amplitudes $C$ and $E$ extracted from Cabibbo-favored charm decays imply
values of $C' = \lambda C$ and $E' = \lambda E$ which may be used in
constructing amplitudes for singly-Cabibbo-suppressed $D^0$ decays involving
$\eta$ and $\eta'$.  In Table \ref{tab:scseta} we write amplitudes multiplied
by factors so that they involve unit coefficient of an
amplitude $SE'$ describing a disconnected ``singlet'' exchange amplitude for
$D^0$ decays \cite{Chiang:2003}.  Similarly the decays $D^+ \to (\pi^+ \eta,
\pi^+ \eta')$ and $D_s^+ \to (K^+ \eta, K^+ \eta')$ may be described in
terms of a disconnected singlet annihilation amplitude $SA'$, written with
unit coefficient in Table \ref{tab:scseta}.  For experimental values we
have used new CLEO measurements as reported in Ref.\ \cite{CLEODs}.  (See
Table \ref{tab:pluseta}.)

\begin{table*}
\caption{Real and imaginary parts of amplitudes for SCS charm decays involving
$\eta$ and $\eta'$, in units of $10^{-7}$ GeV as
predicted in Ref.\ \cite{Chiang:2003}.
\label{tab:scseta}}
\begin{center}
\begin{tabular}{|c|c|r|r|c|} \hline
Amplitude   & Expression & Re & Im & $\ca_{\rm exp}$ \\ \hline
$-\sx \ca(D^0 \to \pi^0 \eta)$ & $2E'-C' +SE'$
     & 0.82 & 9.24 & \\
$-\frac{\st}{2} \ca(D^0 \to \pi^0 \eta\,')$ & $\frac12(C' + E') + SE'$
     & $-2.87$ & 0.56 & \\
$\frac{3}{2 \s} \ca(D^0 \to \eta \eta)$ & $C'+ SE'$
     & $-4.10$ & $-2.33$ & \\
$-\frac{3 \s}{7} \ca(D^0 \to \eta \eta\,')$ & $\frac17(C' + 6E')+SE'$
     & $-1.99$ & 2.63 & \\
$\st \ca(D^+ \to \pi^+ \eta)$ & $T'+2C'+2A'+ SA'$
     & $0.71$ & $-10.68$ & 8.29$\pm$0.38 \\
$-\frac{\sx}{4} \ca(D^+ \to \pi^+ \eta\,')$ & $\frac14(T'-C'+2A')+SA'$
     & 3.25 & $-0.92$ & 4.03$\pm$0.42 \\
$-\st \ca(D_s^+ \to \eta K^+)$ & $-(T'+2C')+SA'$
     & 1.92 & 4.67 & 9.40$\pm$1.05 \\
$\frac{\sx}{4} \ca(D_s^+ \to \eta\,' K^+)$ & $\frac14(2T'+C'+3A')+SA'$
     & 3.10 & $-2.84$ & 3.88$\pm$0.66 \\ \hline
\end{tabular}
\end{center}
\end{table*}

\begin{figure}
\includegraphics[width=80mm]{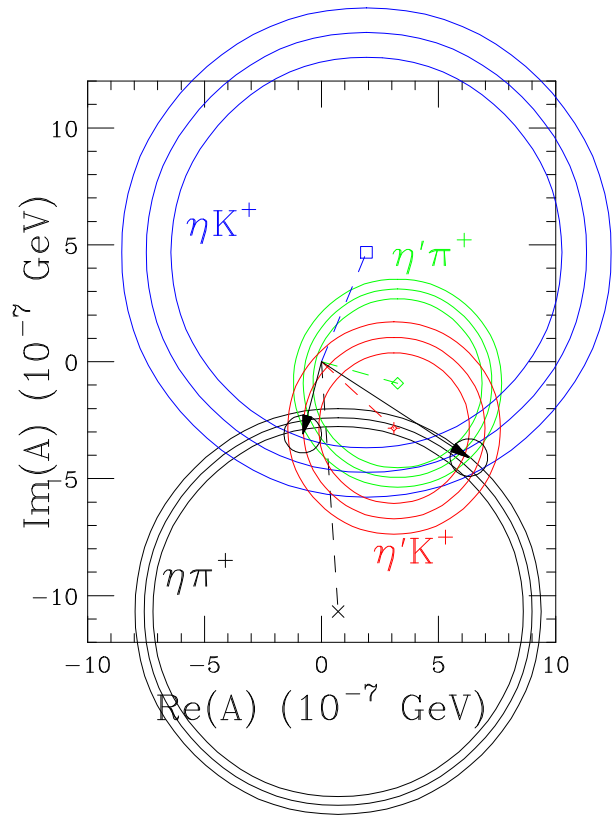}
\caption{Graphical construction to obtain the disconnected singlet annihilation
amplitude $SA'$ from magnitudes of SCS $D^+$ and $D_s^+$ decays involving
$\eta$ and $\eta'$.  Black:  $D^+ \to \eta \pi^+$.  Green: $D^+ \to \eta'
\pi^+$.  Blue: $D_s^+ \to \eta K^+$.  Red: $D_s^+ \to \eta' K^+$.  The small
black circles show the solution regions.
\label{fig:sa}}
\end{figure}

\begin{table}
\caption{Branching ratios and amplitudes for $D^+$ and $D_s^+$ SCS decays
involving $\eta$ and $\eta'$.
\label{tab:pluseta}}
\begin{center}
\begin{tabular}{|c|c|c|c|c|} \hline
Meson   &    Decay     &     $\b$      & $p^*$ &     $\ca$       \\
        &     mode     &  $(10^{-3})$  & (MeV) & ($10^{-7}$ GeV) \\ \hline
$D^+$   & $\pi^+ \eta$ & 3.50$\pm$0.32 &  848  &  4.79$\pm$0.22  \\
        & $\pi^+ \eta'$&  5.3$\pm$1.1  &  681  &  6.58$\pm$0.68  \\
$D_s^+$ &  $K^+ \eta$  & 1.92$\pm$0.43 &  835  &  5.43$\pm$0.61  \\
        &  $K^+ \eta'$ & 2.02$\pm$0.69 &  646  &  6.33$\pm$1.08  \\
\hline
\end{tabular}
\end{center}
\end{table}

We show in Fig.\ \ref{fig:sa} the construction proposed in Ref.\
\cite{Chiang:2003} to obtain the amplitude $SA'$.  Two solutions are found.
In one, $|SA'|$ is uncomfortably large in comparison with the ``connected''
amplitudes, while in the other $|SA'|$ is smaller, but nonzero. 
Corresponding studies of the $D^0$ decays listed in Table \ref{tab:scseta}
\cite{Nisar:2007}, which await further analysis by the CLEO Collaboration, will
permit determination of the corresponding amplitude $SE'$ if one or more
consistent solutions are found.

\section{Doubly-Cabibbo-suppressed decays}

In Table \ref{tab:dcs} we expand amplitudes for doubly-Cabibbo-suppressed
decays in terms of the reduced amplitudes $\ttl \equiv - \tan^2 \theta_C T$,
$\tc \equiv - \tan^2 \theta_C C$, $\te \equiv - \tan^2 \theta_C E$, and
$\ta \equiv - \tan^2 \theta_C A$.

\begin{table*}
\caption{Branching ratios, amplitudes, and representations in terms of
reduced amplitudes for doubly-Cabibbo-suppressed decays.  Amplitudes denoted
by (a) involve interference between the doubly-Cabibbo-suppressed process shown
and the corresponding Cabibbo-favored decay to $\ok + X$.
\label{tab:dcs}}
\begin{center}
\begin{tabular}{|c|c|c|c|c|c|} \hline
Meson &    Decay    &     $\b$      & $p^*$ &    $|\ca|$    & Rep. \\
      &    mode     &  $(10^{-4})$  & (MeV) &($10^{-7}$ GeV)&      \\ \hline
$D^0$ & $K^+ \pi^-$ & 1.45$\pm$0.04 &  861  & 1.54$\pm$0.02 & $\ttl+\te$ \\
      & $K^0 \pi^0$ &      (a)      &  860  &      (a)      & $(\tc-\te)/\s$ \\
      & $K^0 \eta$  &      (a)      &  772  &      (a)      & $\tc/\st$ \\
      & $K^0 \eta'$ &      (a)      &  565  &      (a)      & $-(\tc+3\te)/\sx$
\\ \hline
$D^+$ & $K^0 \pi^+$ &      (a)      &  863  &      (a)      & $\tc+\ta$ \\
      & $K^+ \pi^0$ & 2.28$\pm$0.39 &  864  & 1.21$\pm$0.10 & $(\ttl-\ta)/\s$\\
      & $K^+ \eta$  & 1.01$\pm$0.37 &  776  & 0.85$\pm$0.16 & $-\ttl/\st$ \\
      & $K^+ \eta'$ &   $ < 1.2 $   &  571  &  $ < 1.08 $   & $(\ttl+3\ta)/\sx$
\\ \hline
$D_s^+$& $K^0 K^+$  &      (a)      &  850  &      (a)      & $\ttl+\tc$ \\
\hline
\end{tabular}
\end{center}
\end{table*}

With $\tan \theta_C \simeq 0.23$ one predicts $|\ca(D^0 \to K^+ \pi^-)| = 1.32
\times 10^{-7}$ GeV and $|\ca[D^+ \to K^+(\pi^0,\eta,\eta')] = (0.93,0.83,1.27)
\times 10^{-7}$ GeV, in qualitative agreement with experiment.

\subsection{$D^0 \to (K^0 \pi^0, \ok \pi^0)$ interference}

\begin{figure*}
\mbox{\includegraphics[width=0.46\textwidth]{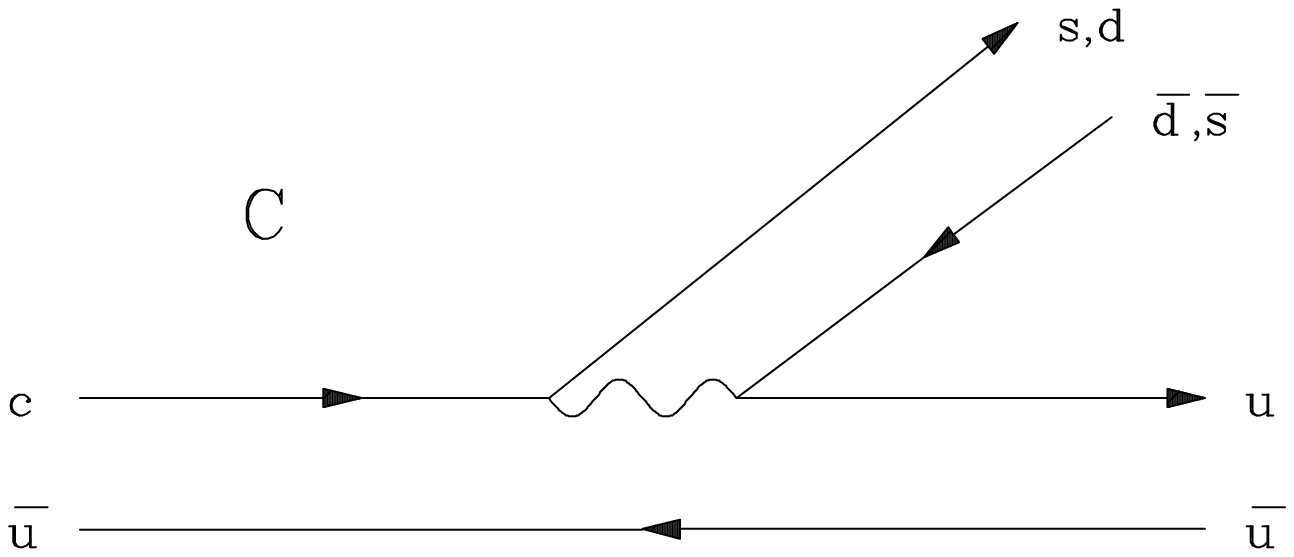} \hskip 0.3in
      \includegraphics[width=0.46\textwidth]{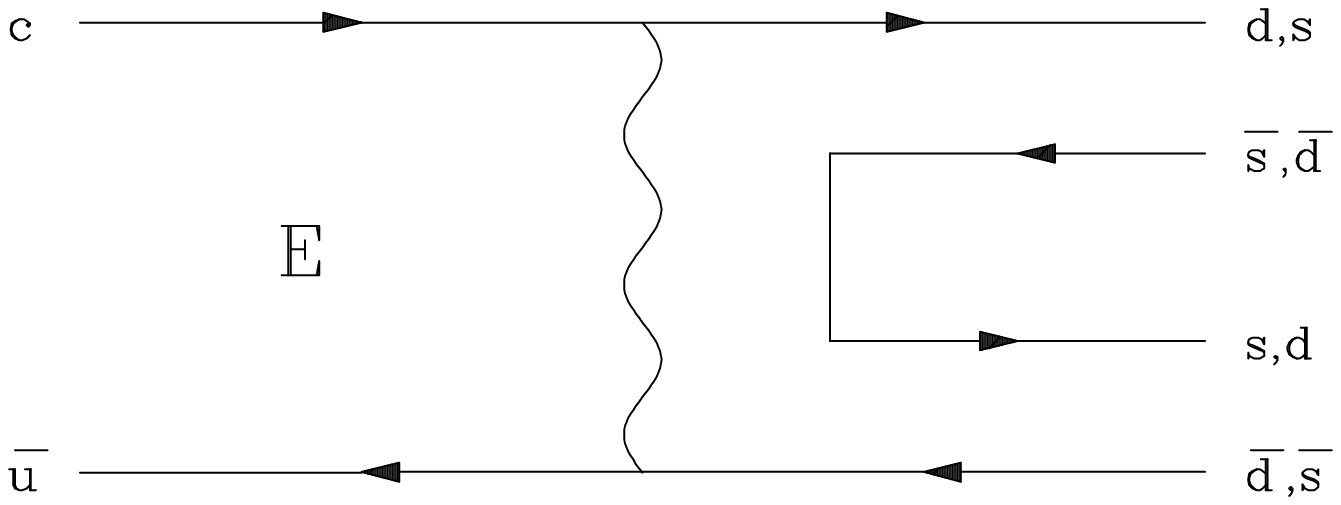}}
\caption{Graphs contributing to $D^0 \to (K^0 \pi^0, \ok \pi^0)$.
\label{fig:Dzint}}
\end{figure*}

The decays $D^0 \to K^0 \pi^0$ and $D^0 \to \ok \pi^0$ are related to
one another by the U-spin interchange $s \leftrightarrow d$, and
SU(3) symmetry breaking is expected to be extremely small in this relation
\cite{Rosner:2006}.  Graphs contributing to these processes are shown in
Fig.\ \ref{fig:Dzint}.

The CLEO Collaboration \cite{CLEODzint} has reported the asymmetry
\beq
R(D^0) \equiv  \frac{\Gamma(D^0 \to K_S \pi^0) - \Gamma(D^0 \to K_L \pi^0)}
 {\Gamma(D^0 \to K_S \pi^0) + \Gamma(D^0 \to K_L \pi^0)}
\eeq
to have the value $R(D^0) = 0.122 \pm 0.024 \pm 0.030$, consistent with
the expected value \cite{Rosner:2006,Bigi} $R(D^0) = 2 \tan^2 \theta_C \simeq
0.108$.  One expects the same $R(D^0)$ if $\pi^0$ is replaced by $\eta$
or $\eta'$ \cite{Rosner:2006}.  Moreover, by similar arguments, one
expects  $A[D^0 \to K^0 (\rho^0,f_0, \ldots)]/
A[D^0 \to \ok (\rho^0,f_0, \ldots)] = - \tan^2 \theta_C$.

\subsection{$D^+ \to (K^0 \pi^+,\ok \pi^+)$ interference}

\begin{figure*}
\mbox{\includegraphics[width=0.31\textwidth]{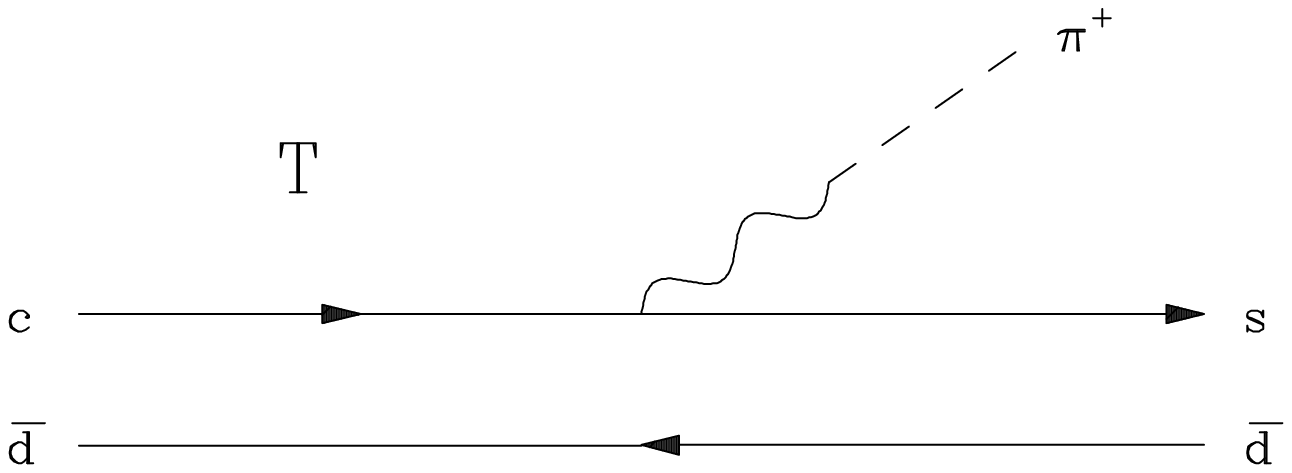} \hskip 0.2in
      \includegraphics[width=0.31\textwidth]{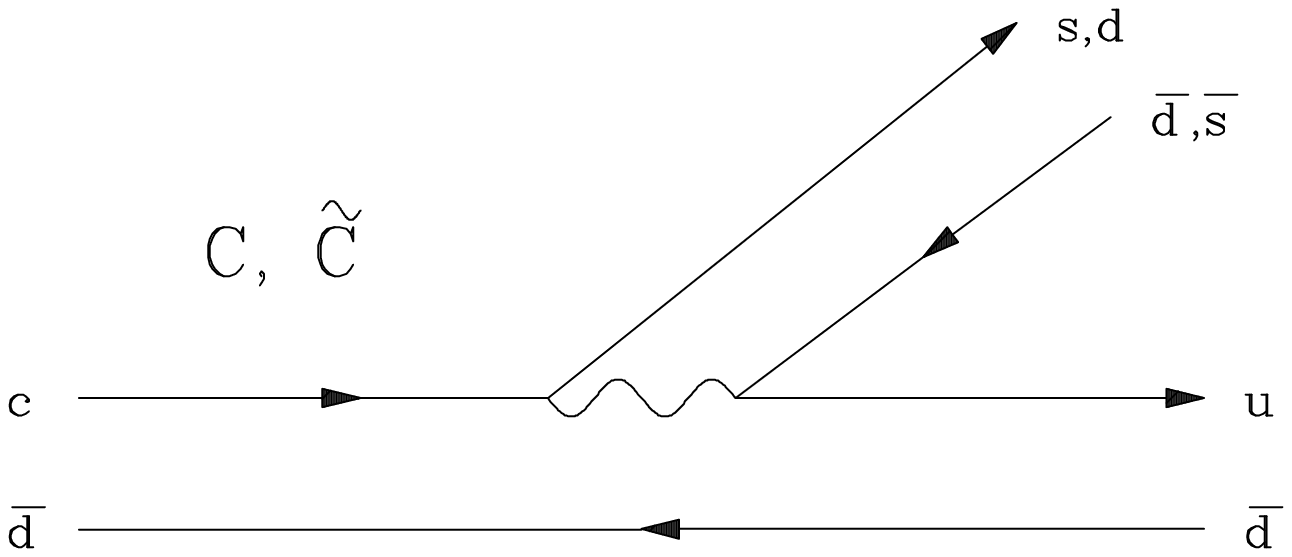} \hskip 0.2in
      \includegraphics[width=0.31\textwidth]{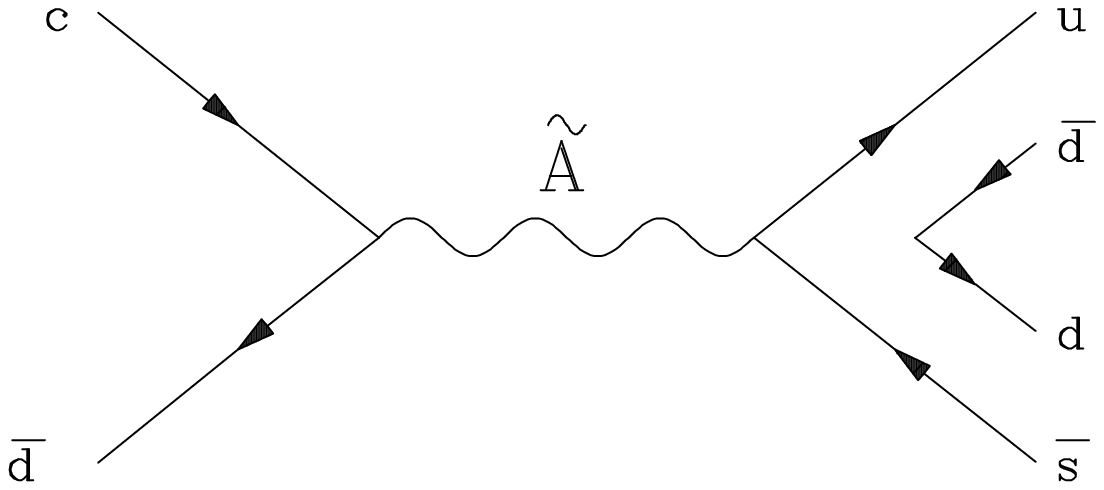}}
\caption{Amplitudes $T$ and $C$ contributing to $D^+ \to \ok \pi^+$;
amplitudes $\tc$ and $\ta$ contributing to $D^+ \to K^0 \pi^+$.
\label{fig:Dpint}}
\end{figure*}

In contrast to the case of $D^0 \to (K^0 \pi^0, \ok \pi^0)$, the decays
$D^+ \to (K^0 \pi^+,\ok \pi^+)$ are not related to one another by
a simple U-spin transformation.  Amplitudes contributing to these
processes are shown in Fig.\ \ref{fig:Dpint}.  Although both processes
receive color-suppressed ($C$ or $\tc$) contributions, the Cabibbo-favored
process receives a color-favored  tree ($T$) contribution, while the
doubly-Cabibbo-suppressed process receives an annihilation ($\ta$)
contribution.  In order to calculate the asymmetry between $K_S$ and
$K_L$ production in these decays due to interference between CF and
DCS amplitudes, one can use the determination of the CF amplitudes 
discussed previously and the relation between them and DCS amplitudes.
Thus, we define
\beq
R(D^+) \equiv \frac{\Gamma(D^+ \to K_S \pi^+) - \Gamma(D^+ \to K_L \pi^+)}
 {\Gamma(D^+ \to K_S \pi^+) + \Gamma(D^+ \to K_L \pi^+)}
\eeq
and predict
\bea
R(D^+) & = & - 2~{\rm Re}~\frac{\tilde{C} + \tilde{A}}{T+C} \nonumber \\
       & = & 2 \tan^2 \theta_C~{\rm Re}~\frac{C+A}{T+C} \nonumber \\
       & = & 0.068 \pm 0.007 ~.
\eea
This is consistent with (though slightly larger in central value than) the
observed value $R(D^+) = 0.026 \pm 0.016 \pm 0.018$ \cite{EHT}.  The
relative phase of $C+A$ and $T+C$ is about $70^\circ$, as can be seen
from Fig.\ \ref{fig:cf}.  The real part of their ratio hence is small.
A similar exercise can be applied to the decays $D_s^+
\to K^+ K^0$ and $D_s^+ \to K^+ \ok$, which are related by U-spin to the
$D^+$ decays discussed here.  The corresponding ratio
\beq
R(D_s^+) \equiv \frac{\Gamma(D_s^+ \to K_S K^+) - \Gamma(D_s^+ \to K_L K^+)}
 {\Gamma(D_s^+ \to K_S K^+) + \Gamma(D_s^+ \to K_L K^+)}
\eeq
is predicted to be
\bea
R(D_s^+) & = & - 2~{\rm Re}~\frac{\tilde{C} + \tilde{T}}{A+C} \nonumber \\
       & = & 2 \tan^2 \theta_C~{\rm Re}~\frac{C+T}{A+C} \nonumber \\
       & = & 0.019 \pm 0.002 ~.
\eea

\section{Strong phases in $(D^0, \od) \to K^- \pi^+$}

The relative strong phase in the CF decay $D^0 \to K^- \pi^+$ and the DCS
decay $\od \to K^- \pi^+$ is of interest in studying $B$ decays involving
neutral $D$ mesons, where these two processes often can interfere.  It was
shown in Ref.\ \cite{Gronau:2001nr} that one could measure this phase by
producing a CP eigenstate $D^0_{CP}$, for example by tagging on a state of
opposite CP at the $\psi(3770)$.  Define decay amplitudes as
\beq
\langle K^- \pi^+ | D^0 \rangle \equiv A e^{i \delta_R}~,~~
\langle K^- \pi^+ | \od \rangle \equiv \bar A e^{i \delta_W}~.
\eeq
The difference $\delta = \delta_R - \delta_W$ of strong phases would
vanish in the SU(3) limit.  At $\psi(3770)$ with $K^- \pi^+$ produced opposite
a state $S_\zeta$ with CP eigenvalue $\zeta$, one would have
\beq
\Gamma(K^-\pi^+, S_\zeta) \approx
A^2 A_{S_\zeta}^2 (1 + 2 \zeta r \cos \delta)~,
\eeq
so by choosing states with $\zeta = \pm 1$ one can measure $(1 + 2 r \cos
\delta) /(1 - 2 r \cos \delta)$, where $r= |\bar A/A|  = 0.057 \simeq \tan^2
\theta_C$.

In an analysis of 281 pb$^{-1}$ of CLEO data \cite{Sun:2007}, the error on
$\cos \delta$ is not yet conclusively determined, as a result of uncertainty
in fits to $D^0$--$\od$ mixing.  For an eventual integrated luminosity at CLEO
of 750 pb$^{-1}$ and a cross section of $\sigma(e^+ e^- \to \psi(3770) \to D
\bar D) = 6$ nb one can estimate by rescaling the calculation in Ref.\
\cite{Gronau:2001nr} an eventual error of $\Delta \cos \delta < 0.2$.

\section{Other theoretical approaches}

One can invoke effects of final state interactions to explain arbitrarily
large SU(3) violations (if, for example, a resonance with SU(3)-violating
couplings dominates a decay such as $D^0 \to \pi^+ \pi^-$ or $D^0 \to K^+
K^-$).  As one example of this approach \cite{Buccella:1996}, both resonant and
nonresonant scattering can account for the observed ratio $\Gamma(D^0 \to
K^+ K^-)/\Gamma(D^0 \to \pi^+ \pi^-) = 2.8 \pm 0.1$.  This same approach
predicted $\b(D^0 \to K^0 \ok) = 9.8 \times 10^{-4}$, a level of SU(3)
violation consistent with the world average of Ref.\ \cite{Yao:2006px} but
far in excess of the recent CLEO value \cite{Ryd:2007}.  The paper of Ref.\
\cite{Buccella:1996} may be consulted for many predictions for $PV$ and
$PS$ final states in charm decays, where $V$ denotes a vector meson and
$S$ denotes a scalar meson.  Results for $PV$ decays also may be found in
Refs.\ \cite{Rosner:1999,Chiang:2003,Chiang:2002,Cheng:2003}.

The recent discussion of Ref.\ \cite{Gao:2007} entails a prediction
$A \simeq - 0.4E$ (recall we were finding $A \simeq -E$), essentially as
a consequence of a Fierz identity and QCD corrections.  Tree amplitudes are
obtained from factorization and semileptonic $D \to \pi$ and $D \to K$ form
factors.  The main source of SU(3) breaking in $\tt/T$ is assumed to come
from $f_K/f_\pi = 1.22$.  Predictions include asymmetries $R(D^{0,+} =
(2 \tan^2 \theta_C,~0.068 \pm 0.007)$, and -- via a sum rule for $D^0
\to K^\mp \pi^\pm$ and $D^+ \to K^+ \pi^0$ -- and expectation of $|\delta|
\simeq 7$--$20^\circ$ (to be compared with 0 in exact SU(3) symmetry).

\section{Summary}

We have shown that the relative magnitudes and phases of amplitudes
contributing to charm decays into two pseudoscalar mesons are describable by
flavor symmetry.  We have verified that there are large relative phases
between the color-favored tree amplitude $T$ and the color-suppressed
amplitude $C$, as well as between $T$ and $E \simeq -A$.

The largest symmetry-breaking effects are visible in singly-Cabibbo-suppressed
(SCS) decays, particularly in the $D^0 \to(\pi^+ \pi^-/K^+ K^-)$ ratio  which
are at least in part understandable through form factor and decay constant
effects.  Decays involving $\eta$, $\eta'$ are mostly describable with
small ``disconnected'' amplitudes, a possible exception being in SCS $D^+$ and
$D_s^+$ decays.

One sees evidence for the expected interference between Cabibbo-favored (CF)
and doubly-Cabibbo-suppressed decays in $D^{0,+} \to K_{S,L} \pi^{0,+}$ decays.
As a result of CLEO's present data on $(D^0,\od) \to K^- \pi^+$, limits are
being placed on the relative strong phase $\delta$ between these amplitudes,
and the full CLEO data sample is expected to result in an error equal to
or better than $\Delta (\cos \delta) = 0.2$.

\bigskip
\begin{acknowledgments}
J.L.R. wishes to thank C.-W. Chiang, M. Gronau, Y. Grossman, and Z. Luo for
enjoyable collaborations on some of the topics described here.  We are
grateful to S. Blusk, H. Mahlke, A. Ryd, and E. Thorndike for helpful
discussions.  This work was supported in part by the United States Department
of Energy through Grant No.\ DE FG02 90ER40560.
\end{acknowledgments}

\bigskip 

\end{document}